%% file: AsteroRedGiants.tex
\documentclass[mypaper,8pt,twoside]{CoAst}
\usepackage{epsf,graphicx,fancyhdr,subfigure}
\input{CoAst_wroclaw-proceedingslayo}

\def\rfr{\smallskip\par\noindent
        \hangindent=7truemm
        \hangafter=1}

\begin{document}
\sf

\chapterCoAst{Asteroseismology of Red Giant stars}%paper title and page heading for even pages
{N.J. Tarrant, W.J. Chaplin, Y.P. Elsworth, S.A. Spreckley, I.R. Stevens} %page heading for odd pages
\Authors{N.J. Tarrant, W.J. Chaplin, Y.P. Elsworth, S.A. Spreckley, I.R. Stevens} 
\Address{
 University of Birmingham, Edgbaston, Birmingham, B15 2TT
}

\noindent
\begin{abstract}
Sun-like oscillations, that is p-modes excited stochastically by convective noise, have now been observed in a number of Red Giant stars. Compared to those seen in the Sun, these modes are of large amplitude and long period, making the oscillations attractive prospects for observation. However, the low Q-factor of these modes, and issues relating to the rising background at low frequencies, present some interesting challenges for identifying modes and determining the related asteroseismic parameters.

We report on the analysis procedure adopted for peak-bagging by our group at Birmingham, and the techniques used to robustly ensure these are not a product of noise. I also show results from a number of giants extracted from multi-year observations with the SMEI instrument. 
\end{abstract}

\Session{ \twoA }
\Objects{Arcturus, $\beta$ UMi}

\section*{Context}

Sun-like oscillations - that is to say p modes stochastically excited by convective noise - have now been observed in a number of K-class Red Giant stars (Tarrant et al., 2007, 2008a, Stello et al. 2008). Being of comparably large amplitude relative to other Sun-like oscillators, these oscillations are readily observable. However, the long time series required for the resolution of individual modes is prohibitive for many instruments. Table \ref{comparison} presents a comparison between the properties of oscillations in the Sun and the range of values expected for a typical K-class giant, based upon the scaling laws of Kjeldsen and Bedding (1995).

In the following sections we discuss the observation of oscillations in K-class giant stars in the context of the study using the SMEI instrument.

\begin{table}[ht!]
\centering

\caption{Comparison of the fundamental parameters of the sun and a K-class giant star.}

\begin{tabular}{ccc}
\hline\hline
	&	Sun		& K-class giant	\\
\hline
T$_{\rm eff}$ (K)		&	5777	&	4000--5000	\\
Luminosity, L$_{\odot}$	&	1 		&	20--400 \\
Radius, R$_{\odot}$		&	1 		&	5--50	\\
Mass, M$_{\odot}$		&	1		&	0.7--6.0	\\
\hline
%Frequency of Max. Power,
$\nu_{\rm max}$, ($\mu$Hz)	&	3050	&	0--40	\\
%Large Spacing,
$\Delta\nu$, ($\mu$Hz)	&	134.9	&	0--6	\\
%Mode Amplitude,
$(\delta L / L)_{\rm 700 nm}$, (ppm)	&	3.7$\pm$0.2	&	100--2000	\\
\hline\hline
\end{tabular}
\label{comparison}
\end{table}

\section*{The SMEI instrument}

The Solar Mass Ejection Imager (SMEI) instrument (Jackson et al., 2004) is aboard the Coriolis satellite. This satellite occupies a Sun-synchronous polar orbit of period $\sim 101$\,minutes, lying along the day-night terminator. Coriolis was launched on 2003 January 6, entering science mode in the spring of that year. The SMEI instrument was designed to detect transient disturbances in the solar wind by means of imaging Compton scattered light from the free electrons in the solar wind plasma. By this means it is possible to map the heliosphere from 0.4 AU to the Earth, and evaluate the usefulness of sensing the heliosphere as a tool for space weather forecasting.

SMEI comprises 3 cameras, each with a field of view of $60 \times 3$ degrees. The cameras are aligned such that the instantaneous total field of view is a strip of sky of size $170 \times 3$ degrees; a near-complete image of the sky is obtained from data on all three cameras after every orbit. Individual images -- which are made from 10 stacked exposures, with a total integration time of about 40\,s -- occupy an arc-shaped $1242 \times 256$-pixel section of each $1272 \times 576$-pixel CCD.  Observations are made in white light, and the spectral response of the cameras is very broad, extending from $\sim 500$ to $\sim 900\,\rm nm$, with a peak at about 700\,nm. Here, we give a brief summary of the main steps of the data analysis pipeline used to generate the light curves

Poor-quality frames having high background are first excluded from any further processing. Processing of the good frames begins with subtraction of bias, calculated from overscan regions at the edges of each frame, and a temperature-scaled dark-current signal. The frames are then flat fielded and spurious signals from cosmic ray hits are removed from the images. The Camera \#2 data suffer from some stray light, and further cleaning of these data is performed to minimize the stray-light contribution (Buffington, private communication). The stray-light problem is concentrated in a small number of pixels on the CCD.

Once the images have been cleaned aperture photometry is performed with a modified version of the \emph{DAOPHOT} routines (Stetsen 1987). The target star is tracked, and its light curve is corrected for the degradation of the CCD over the course of the mission, and a position-dependent correction is applied to compensate for variation of the Point Spread Function (PSF) across the frames. When the star lies within the field of view of one of the cameras, a single photometric measurement of its intensity is therefore obtained once every orbit.

In addition to the work on giant stars reported on later in this report, data from these images has been used in studying the period and amplitude evolution of Polaris (Spreckley \& Stevens, 2008), and in the discovery of new modes of pulsation in gamma Doradus (Tarrant et al., 2008b).

\section*{Issues in analysis}

Analysis of time-series data from SMEI is subject to a number of complications and issues, some of which are universally applicable, and some particular to the instrument itself.

Datasets collected from SMEI show significant artifacts at 1 d$^{-1}$ and higher harmonics (as seen in Figure \ref{daily}). These arise in part as a result of the increased flux of cosmic rays when the satellite passes above the South Atlantic anomaly. An observation will be made once a day while the satellite passes through this region, leading to a higher flux of cosmic rays. As frames with a substantial number of cosmic rays are excluded from the analysis in the pipeline, this introduces a 1 d$^{-1}$ periodicity into the window function. In addition, as the cosmic ray removal process is not perfect, some additional flux from these rays will be present in the photometry, and will contribute a non-sinusoidal flux variation with period one day. A final contribution to the daily artifacts may arise from the stray light issue mention above. 

\figureCoAst{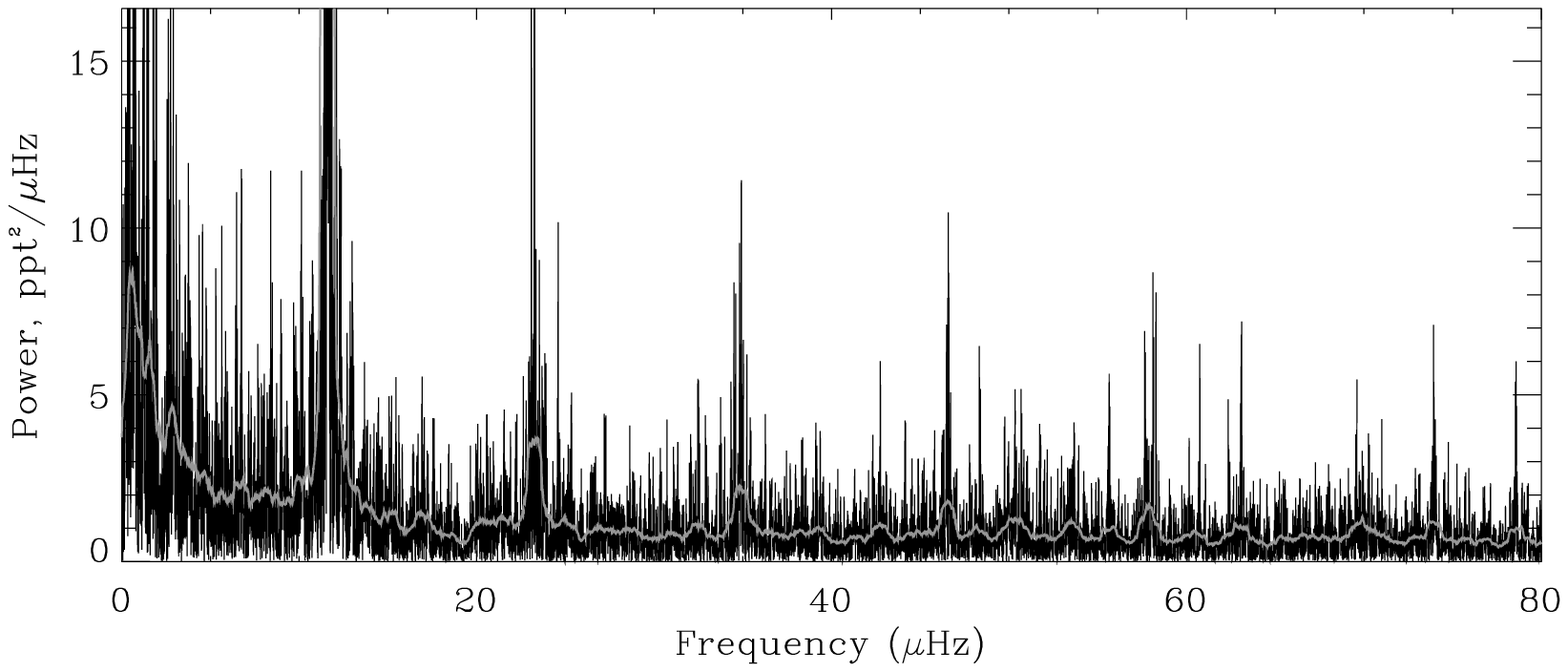}{The SMEI periodogram of $\eta$ Col (HIP 28328, V$_{\rm mag}$=3.9). Daily harmonics are prominent at 11.57 $\mu$Hz and multiples. Here the raw transform is shown in black, and a smoothed spectrum is over plotted in grey.}{daily}{ht}{clip,angle=0,width=90mm}

The strength of the daily harmonics in the periodogram is dependent upon the location of a star - for instance the harmonics are of low amplitude in Polaris - and the brightness, with the features becoming more observable at higher magnitudes. These artifacts at the daily cycles restrict the ability to identify oscillatory modes in the regions surrounding 1 d$^{-1}$ and multiples thereof. Fortunately substantial numbers of K-class giants are expected to have oscillations at frequencies lower than 1 d$^{-1}$, so can be easily differentiated from the the daily harmonics.

In the SMEI data there exists a background which rises steeply at low frequencies, as can be seen in Figure \ref{daily}. This background consists of the intrinsic variability of the star associated with granulation-noise and active regions of the stellar surface, as well as local processes such as instrumental and photon-shot noise. It is natural to expect that other instruments will have similar backgrounds, although the contribution of instrument noise will vary between observatories, and stellar noise manifests differently between observations made in Doppler velocity and photometry.

A large number of the K-class giant stars are predicted to show oscillations below a few $\mu$Hz, that is to say in the region where the background begins to rises rapidly. As the statistical tests (see below) for the significance of a feature depend strongly upon the fitted background within the region of a mode, it is essential to accurately model the background noise. As the excess power associated with the modes will itself raise the apparent background level, the region in which excess power is observed may need to be excluded in determining an accurate model of the background. 

Considering cases where the modes may be resolved, for many stars the large frequency spacings are predicted to be small, of the order of 1 microhertz. This indicated that a lengthy period of observations is required to resolve the individual peaks of a spectrum. 
A final issue arises from the lifetimes of modes in giant stars, as manifested in the width of a peak in frequency space. There appears to be an approximate empirical relationship between the Quality-factor (defined here as the frequency divided by the full width at half maximum [FWHM]), and the period of the oscillations (see Figure \ref{quality}). Considering periods of the giant stars under study, we find that the quality factor for modes can be anticipated to be of the order of 10 to 100, and the FWHM will therefore be of the order of a few tenths of a microhertz. As this width is comparable in size to the expected large frequency separation, we should expect the spectrum to appear very crowded if a significant number of modes are present. In certain cases this may lead to individual modes being unresolvable. The broad FWHM of these oscillations mean power will be spread across multiple bins in a periodogram. This may mean that no single bin achieves a significant signal-to-noise. In this case other statistical tests which consider power in multiple bins within a range, e.g., as described below, may be appropriate.

\figureCoAst{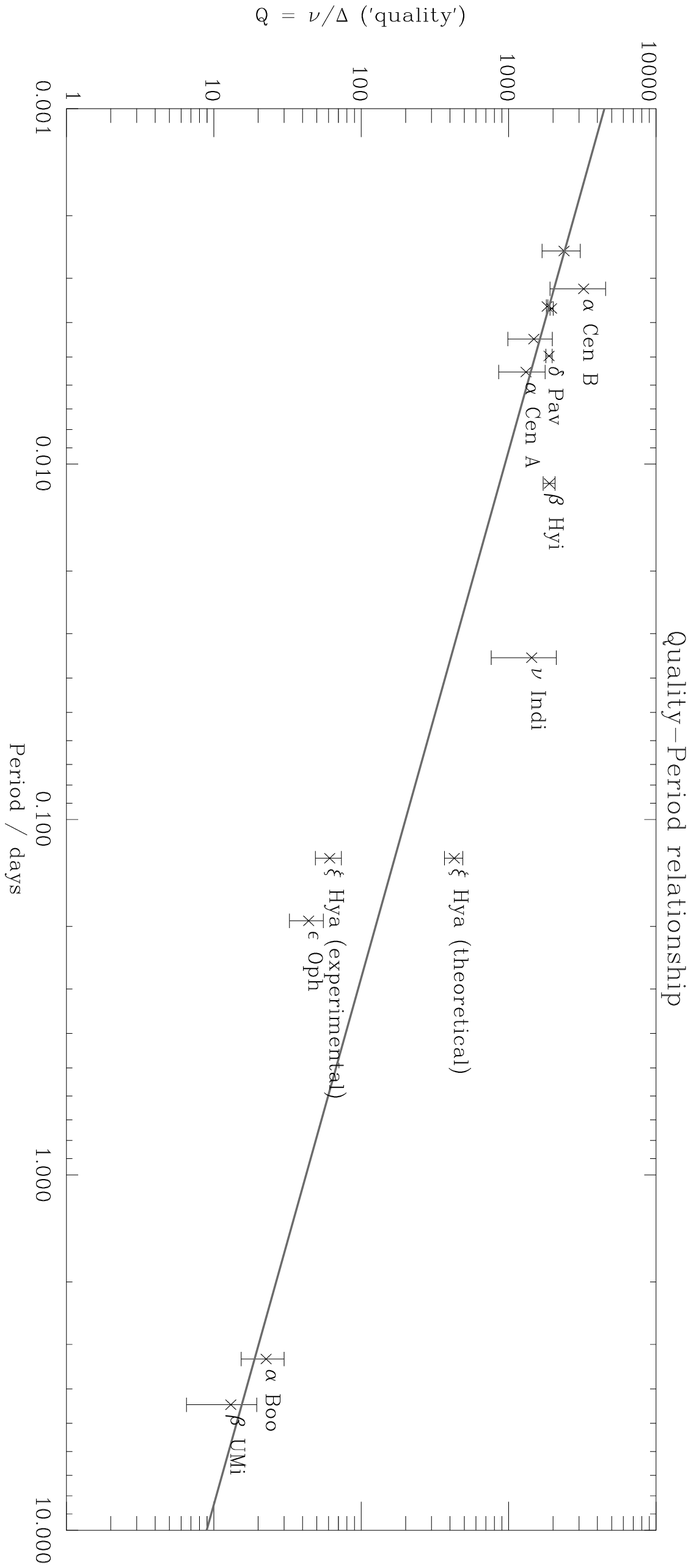}{The apparent quality-period relation seen in Sun-like oscillations. The 
line is a linear fit in log-log space, excluding the theoretical value for $\xi$ Hya. Unlabeled points 
are two determinations for the Sun using GOLF and BiSON data, and determinations at multiple frequencies 
for $\alpha$ Cen A and B. (Using values taken from  Barban et al. (2007), Bedding et al. (2005), Carrier et al (2007),  Houdek and Gough (2002), Kjeldsen et al. (2005) and references therein, Stello et al. (2006), and Tarrant et al. (2007, 2008a))}{quality}{ht}{clip,angle=90,width=85mm}

\section*{Analysis Techniques}

We first wish to identify the region of excess power (i.e., where the modes are located) in the spectrum, in order that the region can be excluded from the fitting to the background and an estimate of the background level, unprejudiced by the excess power, is derived. We first fit an initial background to the periodogram, choosing one of three models for the background:
\begin{enumerate}
\item A power law model of the form $P(\nu) = a + b \nu^{-c}$, reflecting for instance a Brownian noise source.
\item A Harvey model of noise with a memory, consisting of a background and a single component of the form $2 \sigma^2\tau / (1+(2\pi\tau\nu)^2)$, reflecting for instance noise associated with stellar granulation.
\item A Harvey model of noise with memory, consisting of a background and two components, each modelled as in (2), reflecting for instance contributions of both granulation and active regions. 
\end{enumerate}
The best fitting background (as determined by likelihood maximisation) is then compared with a smoothed representation of the data, in which the data has been smoothed using a filter consisting of a Gaussian with a FWHM of twice the expected large frequency spacing. Any excess of power will be thus be visible as a region where the smoothed profile rises above the background power. 

Having thence determined the region in which an excess power is to be found, the background is recomputed with this region excluded from the fitting, again using the three models of the background above. This new background is used to provide the local background level required by the statistical tests described below, and any features which these tests highlight as being significant at a 1\% criterion across a range of tests are noted as possible modes. We fit to these features using the full resonant profile of a mode rather than the Lorentzian approximation, as here the Quality-factor is low and this may have an appreciable effect upon the determined mode parameters.

As a final check, and to determine the errors upon fitted values, which may be underestimated by the formal errors, we run a number of simulations of the data, in which we test whether the effects of the window function can reproduce features in the background comparable in significance to those observed. 

\section*{Statistical Tests}

We analyze data by testing whether features can be explained as being a product of a Gaussian-distributed white-noise background. Under these circumstances the normalised power - that is the power over the mean - in a given bin will be follow a negative exponential distribution (chi-squared with two degrees of freedom). Where the background is not white, periodogram data can be 'whitened' in the frequency domain by dividing throughout by a model of the local background power.

A number of complementary tests can be applied to data to assess whether a given feature shows a significant deviation of power over the background. The most intuitive of these is to look for single spikes which exceed a threshold in normalised power. However in the case of K-class giants where the widths of peaks are anticipated to be broad, one can expect for the peaks to be resolved. This opens the possibility of two further tests of significance - considering the power in two or more adjacent bins, and consideration of whether a significant number of bins rise above a threshold value within a small frequency range. A description of the underlying statistics, and how these tests may be applied to periodogram data can be found in Chaplin et al (2002). 

In a number of circumstances it may be appropriate to consider the sum of power over a number of adjacent bins. This will enable one to detect broad concentrations of power, in which no single notable spike exists and, when considering ranges greater in size to the large-frequency spacing, to determine the envelope in which a power excess associated with the presence of modes exists, even where individual modes within this region may be unresolved. 

\section*{Results}

\begin{figure}[ht]
%\centering

\includegraphics[width=0.5\textwidth]{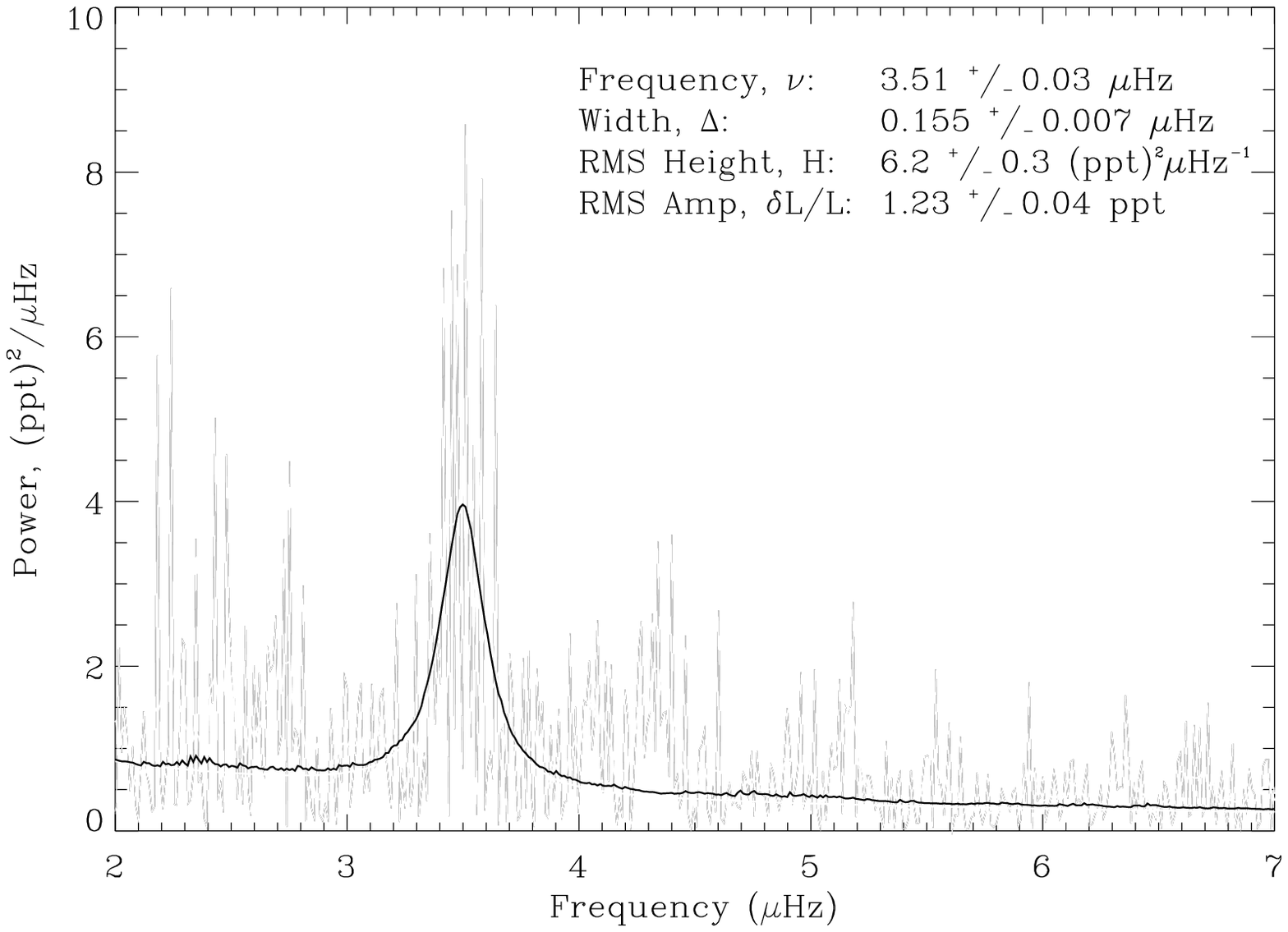}
\includegraphics[width=0.5\textwidth]{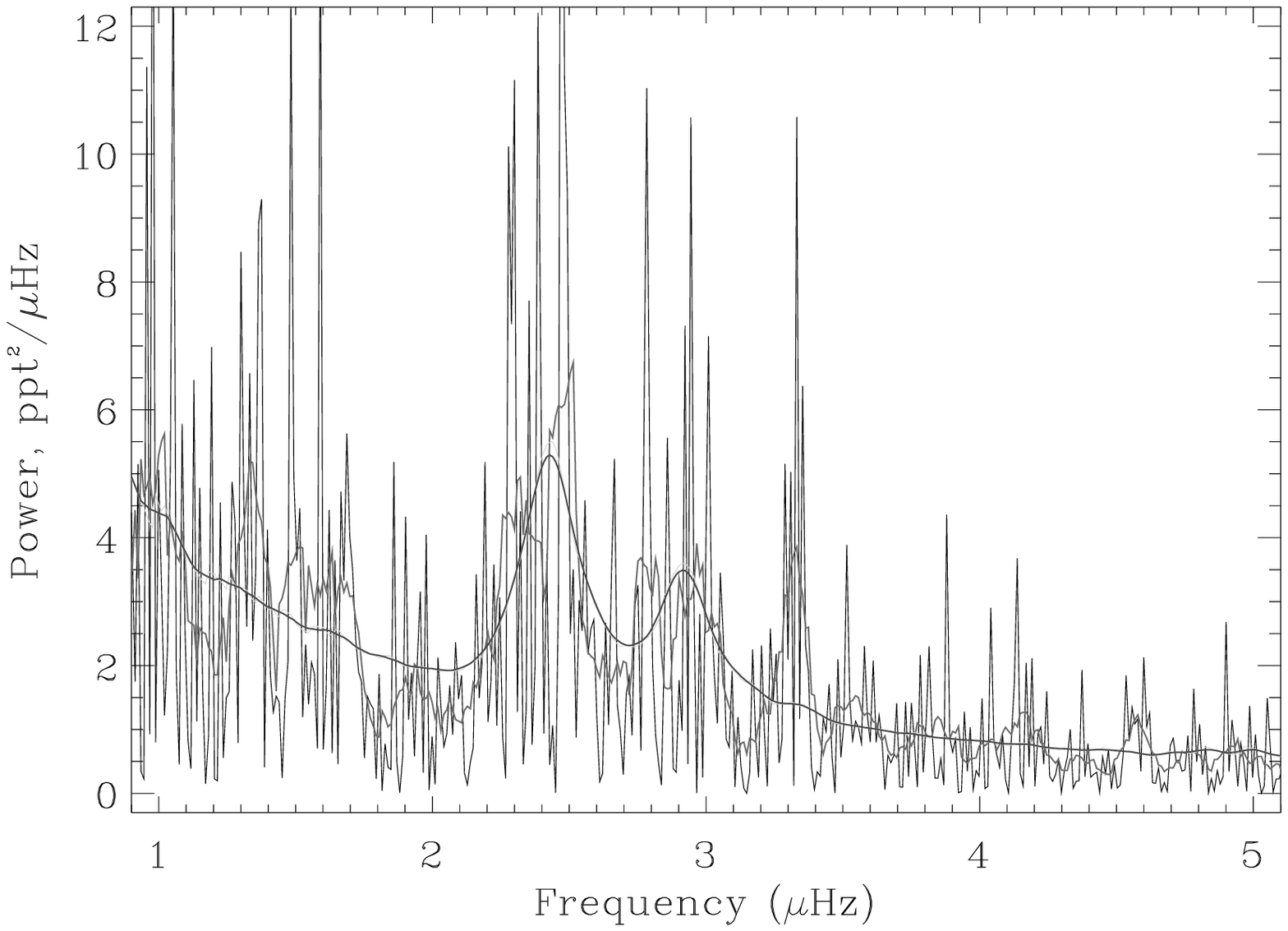}

\caption{A single significant mode, consistent with scaling laws is seen in Arcturus, while $\beta$ UMi seems to show a structure of modes.}

\label{label}

\end{figure}

Using the above procedures we found a single mode in Arcturus (seen in Figure \ref{label}) at a frequency of 3.51 $\mu$Hz, with an RMS amplitude of 1.23 parts-per-thousand (ppt). These values are in reasonable agreement with predictions based upon the scaling laws of Kjeldsen and Bedding (1995), which predict a frequency of $\nu_{\rm max} = 4.7 \pm 1.7$ $\mu$Hz, and an RMS amplitude of $1.3\pm 0.6$ ppt.

In $\beta$ UMi two features at 2.44 and 2.92 $\mu$Hz (seen in Figure \ref{label}, and described in Table \ref{betaumitable}) are identified as being statistically significant. Based upon the assumption that these are modes of adjacent radial order, we have determined that these imply a mass for $\beta$ UMi of 1.3 $\pm$ 0.3 M$_{\odot}$ which differs significantly from evolutionary track and log g mass estimates of $	2.2 \pm 0.3	\ {\rm M}_{\odot } $ and $	2.5 \pm 0.9	\ {\rm M}_{\odot } $  respectively.

\begin{table}

\centering

\caption{ Best-fitting estimates of the identified modes in $\beta$ UMi. }

\label{betaumitable}

\begin{tabular}{cccc}

\hline \hline
Frequency	&	Width (FWHM, $\Delta$)			&	Height				&	RMS amplitude		\\
($\mu$Hz)	&	($\mu$Hz)	&	[(ppt)$^2/\mu$Hz]	&	(ppt)		\\
\hline
	$2.44 \pm 0.04$	&	$0.2 \pm 0.1$	&	$5.4 \pm 2.2$	&	$1.3 \pm 0.4 $	\\
	$2.92 \pm 0.05$	&	$0.2 \pm 0.1$	&	$2.8 \pm 1.1$	&	$0.9 \pm 0.3 $	\\
\hline \hline
\end{tabular}
\end{table}

We are in the process of analyzing further giant stars for the presence of individual modes, having noted excesses of power in a number of stars. 

\section*{Conclusions}

The SMEI satellite is a useful tool for observing oscillations in giant stars, although there are certain issues of which the investigator needs to be aware when approaching the data. The use of rigorous statistical tests is an important tool in confirming the presence of the broad features associated with Sun-like oscillations in K-class giant stars. 

Given the coverage of the whole sky and long time-series available, SMEI has a valuable r\^ole to play in the observation of asteroseismic oscillations, not only amongst the K-class giants highlighted here, but also other variables with periods of the order of a few hours to a few days, for instance gamma Doradus and beta Cepheii pulsators. 

\acknowledgments{
The authors acknowledge the support of STFC. SMEI was designed and constructed by USAFRL, UCSD, Boston College, Boston University, and the University of Birmingham
NJT would personally like to thank the organizers for arranging this interesting workshop, and HELAS for giving him the chance to present his research. 
}

\References{

\rfr Barban, C., Matthews, J. M., de Ridder, J. et al., 2007, A\&A, 468, 1033
\rfr Bedding, T. R., Kiss, L. L., Kjeldsen, H. et al., 2005, MNRAS, 361, 1375
\rfr Carrier, F., Kjeldsen, H., Bedding, T. R. et al., 2007, A\&A, 470, 1059
\rfr Chaplin, W. J., Elsworth, Y., Isaak, G. R. et al., 2002, MNRAS, 336, 979
\rfr Houdek, G. \& Gough, D. O., 2002, MNRAS, 336, 65
\rfr Jackson, B. V., Buffington, A., Hick, P.P. et al., 2004, SoPh, 225, 177
\rfr Kjeldsen, H. \& Bedding, T. R., 1995, A\&A, 293, 87
\rfr Kjeldsen, H., Bedding, T. R, Butler, R. P. et al., 2005, ApJ, 635, 1281
\rfr Spreckley, S. A., Stevens, I. R., 2008, 388, 1239
\rfr Stello, A., Kjeldsen, H., Bedding, T. R. \& Buzasi, D., 2006, A\&A, 448, 709
\rfr Stello, D., Bruntt, H., Preston, H. \& Buzasi, D. 2008, ApJ, 674, 53
\rfr Stetson, P. B, 1987, PASP, 99, 191
\rfr Tarrant, N. J., Chaplin, W. J., Elseworth, Y. et al., 2007, MNRAS, 382, 48
\rfr Tarrant, N. J., Chaplin, W. J., Elseworth, Y. et al., 2008a, A\&A, 483, 43
\rfr Tarrant, N. J., Chaplin, W. J., Elseworth, Y. et al., 2008b, Accepted to A\&A
}

\end{document}

%% file: CoAst_wroclaw-proceedingslayo.tex
\renewcommand{\chaptermark}[1]{\markboth{#1}{}}

\newcommand{\kopf}{\small\itshape Comm. in Asteroseismology \\ Contribution to the Proceedings of the Wroclaw HELAS Workshop, 2008}

\newcommand{\Authors}[1]{\begin{center}\normalsize\bf\sf #1 \end{center}}

\renewcommand{\author}[1]{\begin{center}\normalsize\bf\sf #1 \end{center}}
\newcommand{\Address}[1]{\begin{center}\small\sf #1 \end{center}}

\newcommand{\Session}[1]{{\vspace{3mm}\small \noindent  \hspace*{3mm} Session: } #1 \normalsize}

\newcommand{\Objects}[1]{{\vspace{0mm}\small \noindent  \hspace*{3mm} Individual Objects: } \small #1 \normalsize}

	\newcommand{\twoA}{\small MODES - extracting eigenmode frequencies \newline}

\renewenvironment{abstract}{\section*{Abstract}\normalsize\sf}{}
\newcommand{\References}[1]{\begin{flushleft}{\large References\\}\vspace*{2mm}\small #1 \end{flushleft}}

\newcommand{\chapterCoAst}[2]{\chapter[\sf\normalsize #1\\ \footnotesize \hspace*{5mm}by #2 \sf\normalsize][]{#1\\}\rhead[\fancyplain{}{\sf\footnotesize \center{#1}}]{\fancyplain{}{\sffamily\thepage}}\lhead[\fancyplain{\kopf}{\sffamily\thepage}]{\fancyplain{\kopf}{\sf\footnotesize \center{#2}}}}

\newcommand{\figureCoAst}[5]{\begin{figure}[#4]
\centering
\includegraphics*[#5]{#1}
\caption{#2}
\label{#3}
\end{figure}}

\renewcommand{\chaptername}{}

\newcommand{\acknowledgments}[1]{\vspace*{5mm}\noindent  \textbf{Acknowledgments.} #1}